\documentclass{article} 

\usepackage{iclr2021_conference,times}

\usepackage{microtype}
\usepackage{graphicx}
\usepackage{booktabs}
\usepackage{array}
\usepackage{longtable}
\usepackage{amsmath}
\usepackage{amssymb}
\usepackage{hyperref}
\usepackage{url}
\usepackage[capitalize,noabbrev]{cleveref}

\newcolumntype{L}[1]{>{\raggedright\arraybackslash}p{#1}}

\title{MDArena: Evaluating Coding Agents on Realistic Molecular Dynamics Workflows}

\author{Nithishwer Mouroug Anand\textsuperscript{1}\thanks{These authors contributed equally.}\,\,\thanks{Correspondence: N.M.A. and P.C.B.},
Wei-Tse Hsu\textsuperscript{1}\footnotemark[1],
Kyle Vaccaro\textsuperscript{1,2}, \\
\bfseries Eden James Gage\textsuperscript{1},
Jonathan David Colburn\textsuperscript{1}, Linda Xi Phan \textsuperscript{1}, \\
\bfseries Minjoon Seo\textsuperscript{1}, Kevin Guan\textsuperscript{1}
\& Philip C. Biggin\textsuperscript{1}\footnotemark[2] \\
\normalfont
\textsuperscript{1}Department of Biochemistry, University of Oxford\\
\textsuperscript{2}Scripps Research Institute\\
\texttt{nithishwer.mourouganand@reuben.ox.ac.uk}\\
\texttt{philip.biggin@bioch.ox.ac.uk}
}

\iclrfinalcopy 

\begin{document}

\maketitle

\begin{abstract}
Accelerating scientific discovery is among the most consequential applications of AI, and computational biomolecular simulation stands out as a particularly promising target within this broader effort. Coding agents promise to automate significant portions of this workflow, yet their reliability on realistic molecular dynamics (MD) tasks remains poorly characterized. To address this issue, we introduce \textbf{MDArena}, a benchmark of 50 containerized tasks drawn from active biomolecular simulation projects, spanning 29 molecular systems and 14 broad research protocols, including trajectory analysis, complex system preparation, free-energy protocols, and enhanced sampling. We evaluate six model/harness configurations spanning Codex and OpenCode. Among the evaluated configurations, Codex GPT-5.5 at extra-high reasoning effort performs best, reaching 24/50 Strict-Pass@1 successes (48\%), followed by Codex GPT-5.5 Medium with 21/50, and OpenCode Gemini Flash 3.5 with 20/50. Average correctness and process rewards are substantially higher than strict success rates across all configurations, indicating that agents frequently make meaningful partial progress but fail on the fine-grained details required for reproducible scientific workflows. Hard tasks remain largely unsolved, particularly membrane-protein system preparation and alchemical free-energy setup, both unsolved or near-unsolved by every evaluated configuration. MDArena thus exposes a substantial gap between the usefulness of coding agents as supervised assistants and their reliability as autonomous MD researchers, while providing a reproducible and extensible platform for tracking progress toward closing it.

\end{abstract}

\section{Introduction}
Molecular dynamics (MD) provides a computational microscope for studying the motion of molecular systems. Over the past decades, it has been widely used across scientific disciplines and delivered insights into molecular mechanisms, dynamics, and functions in numerous studies ~\citep{hollingsworth2018molecular, huggins2019biomolecular}. While its underlying theory is well-established, carrying out a scientifically defensible MD workflow remains technically demanding, as it often requires meticulous system preparation, parameterization, selection of appropriate sampling protocols, and analysis tailored to the scientific question. These activities require substantial expertise and are often repeated across systems and projects, creating a strong incentive to automate their execution.

Recent advances in large language models (LLMs) and coding agents offer a promising route toward such automation ~\citep{zhao2023survey, liu2024large}. In particular, MD is well-suited to agentic assistance because its workflows are mature, modular, and well-supported by a variety of simulation and analysis software ~\citep{abraham2015gromacs,eastman2023openmm,michaud2011mdanalysis}. Equipped with language-based reasoning and the capability to directly interact with computational environments, an agent could in principle translate a scientific objective into an actionable workflow, orchestrate available tools, and execute iteratively adapting protocols given intermediate results or failures. Still, reliable automation of MD workflows and its own evaluation are non-trivial. The challenge lies in the sequential coupling between workflow stages: an error introduced at a certain step, such as an incorrect protonation state or ligand topology, can propagate and invalidate subsequent calculations. Compounding this problem, many mistakes are silent, as a successfully executed simulation may still produce plausible results even when the underlying system or protocol is flawed. A benchmark must therefore assess not only whether an agent can execute MD software successfully, but also whether its workflow and outputs are scientifically valid, especially in realistic scenarios that mirror actual research projects or drug-discovery campaigns. Existing benchmarks and evaluations often fall short on this front ~\citep{kumar2026mdgym, campbell2026mdcrow}, relying on isolated toy problems and narrow research contexts while providing limited support to evaluate the validity of an agent's workflow.

To address this need, we introduce \textbf{MDArena}, an executable benchmark for evaluating coding agents on realistic molecular-simulation workflows. The initial benchmark contains 50 expert-curated tasks derived from real research projects, each accompanied by a containerized environment and formulated as a bounded scientific problem with a clearly defined objective and verifiable outcome. Characterized by three difficulty levels, the benchmark tasks span diverse biomolecular systems and simulation software, covering 14 method classes that include membrane-protein preparation, parameterization of non-standard residues/molecules, enhanced-sampling protocols, quantum-chemistry workflows, and advanced data analysis.

For evaluation, MDArena reports two scores using a two-part verifier: a correctness score, reported by deterministic checks that examine the properties of resulting artifacts, and a process score, assigned by an LLM-as-judge that scrutinizes the agent's execution trajectory against predefined criteria covering scientific decision-making and result interpretation. Reported separately, these two complementary scores distinguish valid solutions from superficially convincing workflows and reveal cases in which a sound process nevertheless produces an incorrect result. 

Using MDArena, we evaluate six contemporary model--harness configurations through Harbor ~\citep{Harbor_Framework} under a common execution protocol and two-part verification scheme. We find that the evaluated agents can solve many well-specified tasks, but remain brittle on long-horizon workflows. Even the strongest configuration solves fewer than half of the tasks, with membrane-protein preparation and other complex workflows remaining largely unsolved. Across models, agents frequently produce plausible intermediate artifacts and make substantial progress, yet overlook details that prevent a fully correct solution. 

These results characterize the frontier of agentic assistance in MD at the time of evaluation: the strongest evaluated coding agents are already useful for bounded tasks under supervision, but cannot yet be relied upon to conduct complete research workflows autonomously.


\section{Related works}


The growing ubiquity of AI agents in scientific computing has driven a parallel effort to benchmark them on realistic scientific workflows, and molecular dynamics is no exception. Yet most existing benchmarks remain narrow in scope and fail to capture the breadth of high-level tasks relevant to MD workflows.

MDCrow ~\citep{campbell2026mdcrow} was the first to evaluate LLM agents with 40+ tools on 25 MD tasks. However, this benchmark is limited in scope, and modern coding agents now nearly saturate this suite. This was followed by NAMD-Agent ~\citep{chandrasekhar2025namdagent} which automates the CHARMM-GUI interface ~\citep{jo2008charmm} via Gemini ~\citep{team2023gemini} and Selenium. NAMD-Agent primarily serves as a system demonstration and does not introduce an accompanying benchmark. ChemGraph ~\citep{pham2026chemgraph} applies multi-agent orchestration to computational chemistry workflows, including geometry optimization, reaction enthalpy, and Gibbs free energy. It reports 100\% accuracy on all tasks tested, although the task suite is relatively small and focuses on textbook-scale problems. DynaMate ~\citep{guilbert2025dynamate} was the first to target biomolecular MD with realistic tasks, evaluating a 12-system benchmark and automating the industry-relevant MM/GBSA binding-affinity estimation, though it remains a system demonstration rather than a held-out benchmark. MDAgent2 ~\citep{shi2026mdagent2} contributes MD-EvalBench, the first dedicated MD benchmark suite, alongside a domain-adapted Qwen3-8B model \citep{yang2025qwen3} trained via continued pretraining, supervised fine-tuning, and reinforcement learning with execution as the reward signal. Its scope, however, is restricted to LAMMPS ~\citep{thompson2022lammps} code generation, with no biomolecular tasks. PolyJarvis ~\citep{zhao2026polyjarvis} extends this direction to polymer MD, orchestrating polymer MD toolkits via Model Context Protocol (MCP) servers, but remains restricted to polymer MD.

MDGym ~\citep{kumar2026mdgym} is a dedicated benchmark in this line, evaluating agents on GROMACS ~\citep{abraham2015gromacs} and LAMMPS ~\citep{thompson2022lammps} workflows rather than merely demonstrating agent capability. MDGym provides an important foundation, while also leaving several opportunities for broader evaluation. MDGym's current biomolecular tasks score agents against simulation output directly, which can conflate protocol correctness with sampling noise. The task suite currently focuses on simulation execution given pre-built structures and topologies, leaving system preparation and parameterization as a promising direction for future extension. It also concentrates on well-known soluble protein systems, which leaves room to broaden coverage in follow-up work. Additionally, tasks are framed as property prediction problems with fixed accuracy thresholds (e.g., 5\% deviation), an approach that could be complemented by grading strategies better suited to settings without a well-defined ground truth. The reported failure-mode taxonomy provide a useful starting point that could be complemented by more fine-grained, case-level analysis in future iterations.

MDArena departs from these works along three axes. First, provenance: tasks are drawn from active research projects in a working MD group rather than authored as benchmark artifacts, allowing continual addition of novel research workflows as they arise. Second, scope: coverage extends beyond prior work into alchemical free-energy perturbation (FEP) ~\citep{mey2020best} workflows, membrane-protein systems, protein-ligand complexes, and enhanced-sampling methods such as replica exchange of expanded ensembles (REXEE)~\citep{hsu2024replica} and alchemical metadynamics~\citep{hsu2023alchemical}. Third, evaluation: agent outputs and execution trajectories are graded against a dual-pronged scheme combining human-authored, criterion-anchored rubrics with deterministic correctness checks designed to catch subtle but common agent errors that rubric grading alone can miss. Together, these yield a more robust and diagnostically informative evaluation than execution- or output-matching alone.

\section{Benchmark Construction}

\begin{figure}[tbp]
\centering
\includegraphics[width=\textwidth]{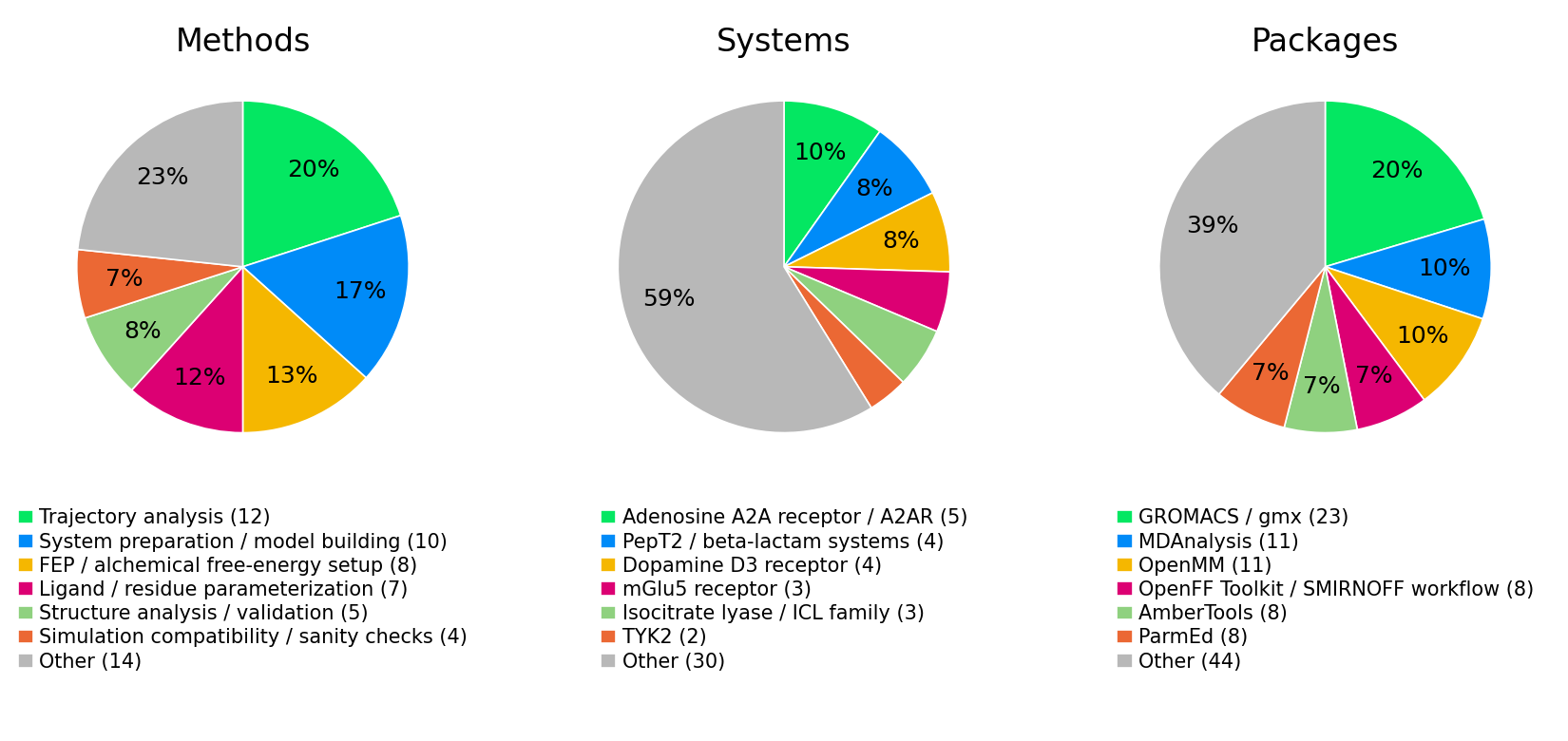}
\caption{Composition of the MDArena dataset. Pie charts showing the distribution of tasks across three axes: method class, system type, and software package.}
\label{fig:dataset-composition}
\end{figure}

MDArena was designed around three goals. Realism: tasks are drawn from authentic MD workflows used in active research projects rather than toy examples. Diversity: tasks span a broad range of molecular systems, methods, analysis routines, and simulation packages (see \cref{fig:dataset-composition}). Robustness: verifiers are built to capture a broad range of agent failure modes, with tasks and rubrics specifically designed to catch subtle errors. 

\subsection{Dataset construction and curation}

An initial collection round yielded more than 100 candidate MD tasks from group members. The associated systems, trajectories, and intermediate files were drawn from completed or ongoing research projects ~\citep{hsu2022identifying, joshi2026structural}, grounding the benchmark in realistic scientific workflows. Candidate problems were then filtered against two criteria: alignment with the dataset's design goals, and difficulty sufficient to remain challenging for modern coding agents. This narrowed the pool to 50 high-quality tasks.
The filtering criteria were chosen to probe the specific factors limiting AI agents in autonomously conducting MD-based biomolecular research. These include: negative examples, where the requested action is impossible or scientifically inappropriate, to probe judgement rather than execution alone; long-horizon tasks, to stress-test context-length limitations; and underspecified tasks, which lack full problem detail, to evaluate an agent's ability to infer research intent from realistic but incompletely specified descriptions.

The final dataset spans 29 systems, 22 software packages and toolkits, and 14 broad method classes (see \cref{tab:task-metadata} for per-task tags). The tasks exercise a broad range of established scientific software, spanning simulation and trajectory analysis (GROMACS~\citep{abraham2015gromacs}, OpenMM~\citep{eastman2023openmm}, MDAnalysis~\citep{michaud2011mdanalysis}), system preparation and small-molecule parameterization (AmberTools~\citep{case2023ambertools}, ACPYPE~\citep{sousa2012acpype}, the OpenFF Toolkit and Sage small-molecule force field~\citep{boothroyd2023development}, RDKit~\citep{landrum2013rdkit}, Open Babel~\citep{o2011open}), free-energy and enhanced sampling workflows (OpenFE~\citep{zhang2023openfe}, alchemlyb~\citep{wu2024alchemlyb}, PLUMED~\citep{tribello2014plumed}), quantum chemistry (ORCA~\citep{neese2020orca}), and validation (physical\_validation~\citep{merz2018testing,merz2022physical_validation}). Each task was assigned an easy, medium, or hard difficulty label by the task author and subsequently reviewed, based on factors including workflow length, required domain knowledge, and technical complexity for modern coding agents. The benchmark set comprises 11 easy, 23 medium, and 16 hard tasks.

\subsection{Verifier Design}

MD workflows, like many computational science workflows, are difficult to verify reliably, largely due to the absence of a ground truth and the inherently stochastic nature of simulation ~\citep{wan2021uncertainty,communicationsbiology2023checklist}. MDArena therefore focuses on the stages surrounding simulation execution: preprocessing and system setup on one end, and analysis and post-processing on the other. Importantly, these are in fact the stages where much of the technical difficulty lies and where outputs can often be evaluated deterministically. Once a valid simulation input, such as a GROMACS TPR file, has been generated, launching the calculation is comparatively straightforward. 

To maximize verifier coverage without imposing overly rigid constraints on agents, we adopted a two-pronged verification scheme combining deterministic correctness checks with LLM-as-judge process checks. The deterministic correctness checks confirm that expected outputs are present and valid, and satisfy numerical, structural, or software-level requirements. The associated correctness reward is defined as the fraction of deterministic criteria satisfied. The process checks instead assess whether the broader, goal-oriented steps taken by the agent are reasonable, without penalizing agents that reach the correct answer through an unorthodox route. These checks are kept minimal, covering only the steps strictly necessary to reach a correct answer given the task instructions. Across the benchmark, the process verifier contains 243 author-defined criteria, with a median of four per task and a range of one to sixteen. Most are graded as binary pass/fail criteria, while a few use a Likert scale. 

\subsection{Implementation framework}

To maximize adoption, MDArena is implemented in the Harbor framework ~\citep{Harbor_Framework}, an agent-evaluation framework from the makers of Terminal-Bench ~\citep{terminalbench2025}, making it straightforward for researchers both to evaluate their own agents against the benchmark and to extend MDArena by adding new self-contained tasks. Each Harbor task is paired with a dedicated Docker image, giving every task an isolated, reproducible environment ~\citep{moreau2023containers}. For every hard task and most medium tasks (31 of 50 in total), we additionally provide a reference solution: an executable script that Harbor's built-in Oracle agent runs in place of an agent ~\citep{Harbor_Framework}. Each reference solution is required to pass all correctness and process checks (reward 1), which serves two purposes --- it confirms that the task is well-posed and solvable, and it validates the verifier itself, ensuring that a correct workflow is scored as a full success rather than penalized by an overly strict check. LLM-as-judge evaluation uses Claude Opus 4.8 as the judge across all tasks, chosen to ensure consistent grading. For each task, the judge is given the agent's full execution trajectory (trajectory.json) together with a task-specific set of output files, and scores the task's process criteria under a 300s to 600s judge timeout. 

\section{Experimental Setup}

\subsection{Harness and model selection}
A coding agent consists of a base language model paired with a harness, the software layer that provides tools, controls execution, and mediates access to the task environment.
Recent research suggests that harness design can substantially influence coding-agent performance alongside the capabilities of the underlying model  \citep{yang2024sweagent,wang2025openhands,lee2026metaharness,lin2026harnessbenefit,lin2026agenticharness}. To examine the contribution of both model and harness choice, we evaluated agents using two distinct harnesses: OpenCode, an open-source harness, and Codex, OpenAI's proprietary closed-source harness. Within OpenCode, we tested four models spanning proprietary and open-weight families and a range of inference costs: Gemini-3.5-Flash and Gemini-3.1-Pro-Preview (both proprietary), and DeepSeek-V3.2-MAAS ~\citep{liu2024deepseek} and Qwen3-235B-A22B-Instruct-2507-MAAS ~\citep{yang2025qwen3} (both open-weight). Within Codex, we evaluated GPT-5.5 at medium and extra-high (XHigh) reasoning effort to probe frontier-model performance. Each model--harness pair was evaluated once per task, yielding a single-run Pass@1 estimate.

\subsection{Execution environments and runtime limits}

Each task in MDArena was implemented in a Docker container with allocated 2--4 CPUs, 8 GB RAM, 10 GB storage, no GPUs, and internet access enabled. By default, tasks use a 900-second agent wall-clock timeout and a 900-second verifier timeout. We treat the agent timeout as part of the benchmark protocol rather than as an infrastructure specification, as it directly affects reported performance. A uniform 900-second budget disadvantaged larger, reasoning-heavy ``thinking'' models at hard tasks, which often required more time to plan and execute before producing an answer. To prevent runtime budget from confounding task performance, we extended the agent timeout for select computationally intensive tasks. Verifier timeouts were similarly extended for specific tasks. Timeout and other runtime failures were counted as unsuccessful tasks for Strict-Pass@1, but are reported separately from verifier-scored failures to distinguish operational errors from incorrect completed workflows.

\section{Results}

\subsection{Overall Accuracy}

\begin{figure}[!ht]
\centering
\includegraphics[width=\textwidth]{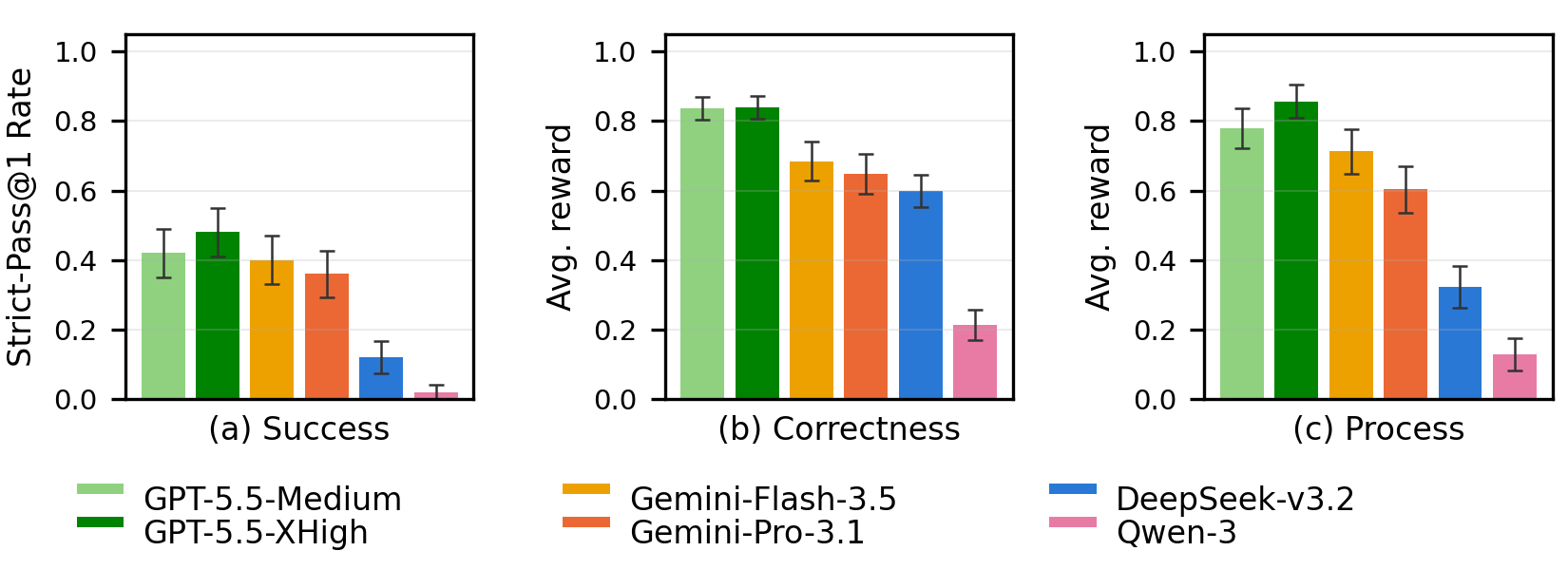}
\caption{Overall agent performance across all MDArena tasks. (a) Strict-Pass@1: a binary metric scored as success only if all process and correctness checks for a task pass. (b) Average correctness reward and (c) average process reward: each task defines multiple correctness and process checks, and we report the mean pass rate across these checks, averaged over all MDArena tasks. Error bars show standard errors across tasks, using binomial standard errors for pass/fail success rates}
\label{fig:success-reward-triptych}
\end{figure}

The overall performance on MDArena is summarized in \cref{fig:success-reward-triptych} and \cref{tab:model-summary}. Full success is measured by Strict-Pass@1, a binary metric requiring all process and correctness checks to pass, while partial credit is captured separately via average correctness and process rewards. Strict-Pass@1 rates are modest across the board: the best-performing configuration, Codex GPT-5.5 XHigh, solves 24/50 tasks (48\%), followed by Codex GPT-5.5 Medium at 21/50 (42\%), OpenCode with Gemini Flash 3.5 at 20/50 (40\%), and Gemini Pro 3.1 at 18/50 (36\%). Open-weight models trail substantially, with DeepSeek-V3.2-MAAS solving 6/50 tasks (12\%) and Qwen3-235B-A22B-Instruct-2507-MAAS solving just 1/50 (2\%).

The gap between Strict-Pass@1 and the average reward metrics indicates that agents often possess strong underlying capability but fall just short of a fully correct solution in these multi-step tasks. GPT-5.5 Medium and XHigh achieve similar average correctness rewards (0.836 and 0.838, respectively), but XHigh attains a notably higher average process reward (0.856 versus 0.778), consistent with its three additional strict successes. This suggests the higher-effort model is more deliberate in its execution, accounting for edge cases that are predominantly captured by the process checks. Gemini-3.5-Flash and Gemini-3.1-Pro performed similarly in terms of the Strict-Pass@1 rate and average correctness reward, though Gemini-3.5-Flash slightly outperformed Gemini-3.1-Pro in the average process reward. DeepSeek shows a similar dissociation at lower absolute performance: a moderate average correctness reward (0.599) alongside a much lower process reward (0.323), suggesting it often produces locally plausible artifacts that fail to satisfy the broader scientific workflow. Qwen, by contrast, frequently fails before even reaching the substantive stages of the workflow, leaving little or no scientifically usable output.

\begin{table}[tbp]
\caption{Overall baseline performance on the 50-task MDArena benchmark. Cost is the mean recorded API cost per task.}
\label{tab:model-summary}
\centering
\small
\begin{tabular}{lrrrr}
\toprule
Agent & Strict-Pass@1 & Correctness & Process & Cost \\
\midrule
GPT-5.5 XHigh & 24/50 (48\%) & 0.838 & 0.856 & USD 2.10 \\
GPT-5.5 Medium & 21/50 (42\%) & 0.836 & 0.778 & USD 0.96 \\
Gemini Flash 3.5 & 20/50 (40\%) & 0.684 & 0.713 & USD 1.55 \\
Gemini Pro 3.1 & 18/50 (36\%) & 0.648 & 0.604 & USD 1.25 \\
DeepSeek-V3.2 & 6/50 (12\%) & 0.599 & 0.323 & USD 0.28 \\
Qwen3-235B & 1/50 (2\%) & 0.214 & 0.129 & USD 0.13 \\
\bottomrule
\end{tabular}
\end{table}

\subsection{Costs and Efficiency}

\begin{figure}[!ht]
\centering
\includegraphics[width=\textwidth]{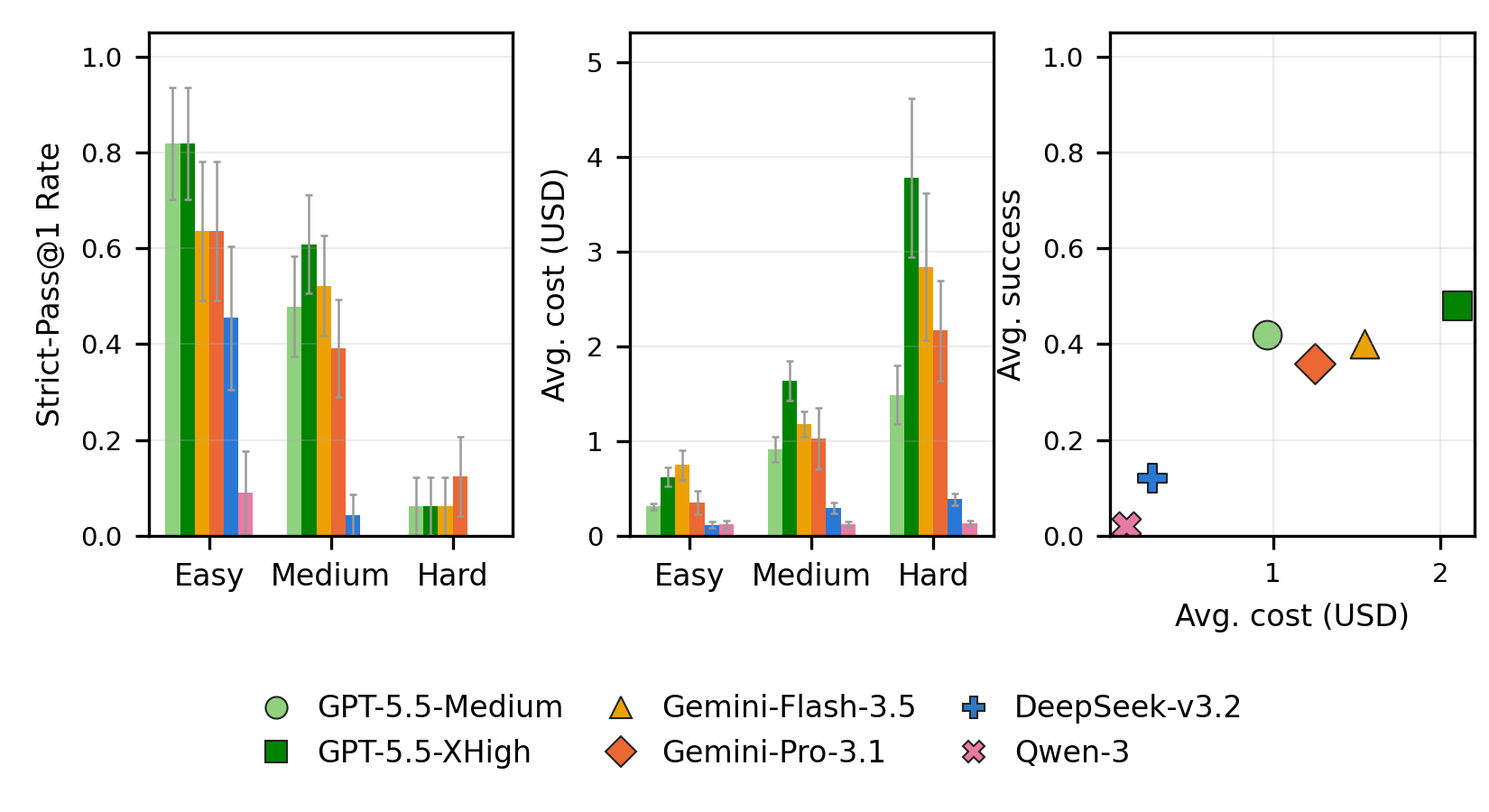}
\caption{Model performance and cost across task difficulty. (a) Strict-Pass@1 success rate by difficulty. (b) Average cost per task by difficulty: mean API cost per task for each model, split across easy, medium, and hard MDArena tasks. (c) Strict-Pass@1 versus average recorded API cost across model–harness configurations.}
\label{fig:cost-by-difficulty}
\end{figure}

Performance and cost breakdown by difficulty is summarized in \cref{fig:cost-by-difficulty}.  The benchmark contains 11 easy, 23 medium, and 16 hard tasks. All agents perform reasonably on easy tasks relative to their overall scores, but hard tasks remain largely unsolved. Both GPT-5.5 configurations solve 9/11 easy tasks and just 1/16 hard tasks; XHigh's overall advantage comes mainly from medium tasks, where it solves 14/23 compared with Medium's 11/23. A similar pattern holds for Gemini: both configurations solve the same 7 easy tasks, but Gemini Flash 3.5 solves more medium tasks than Gemini Pro 3.1 (12/23 versus Pro's 9/23), while Pro reaches slightly more hard-task successes (2/16 versus Flash's 1/16). Gemini 3.5 Flash’s stronger overall performance than Gemini 3.1 Pro is not unexpected, as Flash is positioned specifically for coding, tool use, and long-horizon agentic workflows, whereas Pro is positioned more strongly around advanced reasoning and complex problem solving. What is notable is that Pro’s reasoning-oriented profile does not confer an advantage on MDArena. Its higher action count and more iterative behavior suggest that sustained tool use and execution may matter more than reasoning strength alone for these workflows.

Cost rises with difficulty for every configuration, tracking increases in both time on task and token usage. GPT-5.5 XHigh costs roughly USD 2.10 per task on average, more than twice GPT-5.5 Medium's USD 0.96, for a six-percentage-point gain in Strict-Pass@1. Gemini Flash 3.5 is likewise more expensive than Gemini Pro 3.1 (USD 1.55 versus USD 1.25 per task) but achieves higher success, correctness, and process rewards; this follows from Flash taking more actions per task and behaving more agentically overall, which may contribute to both its higher cost and stronger performance. This cost-performance reversal is a reminder that cheaper or nominally stronger base models do not necessarily yield better agentic performance once harness behavior, tool use, and failure recovery are taken into account.

Success generally improves with cost (\cref{fig:cost-by-difficulty}), as expected, but GPT-5.5 Medium occupies a favorable cost-performance position among the evaluated configurations: within the Codex harness, it costs roughly 23\% less than Gemini Pro 3.1 and 38\% less than Gemini Flash 3.5 under OpenCode, while still achieving higher success on both counts, suggesting that the complete model--harness configuration, not just the base model alone, shapes where a configuration lands on this frontier.

\subsection{Performance across task categories}

\begin{figure}[!ht]
\centering
\includegraphics[width=\textwidth]{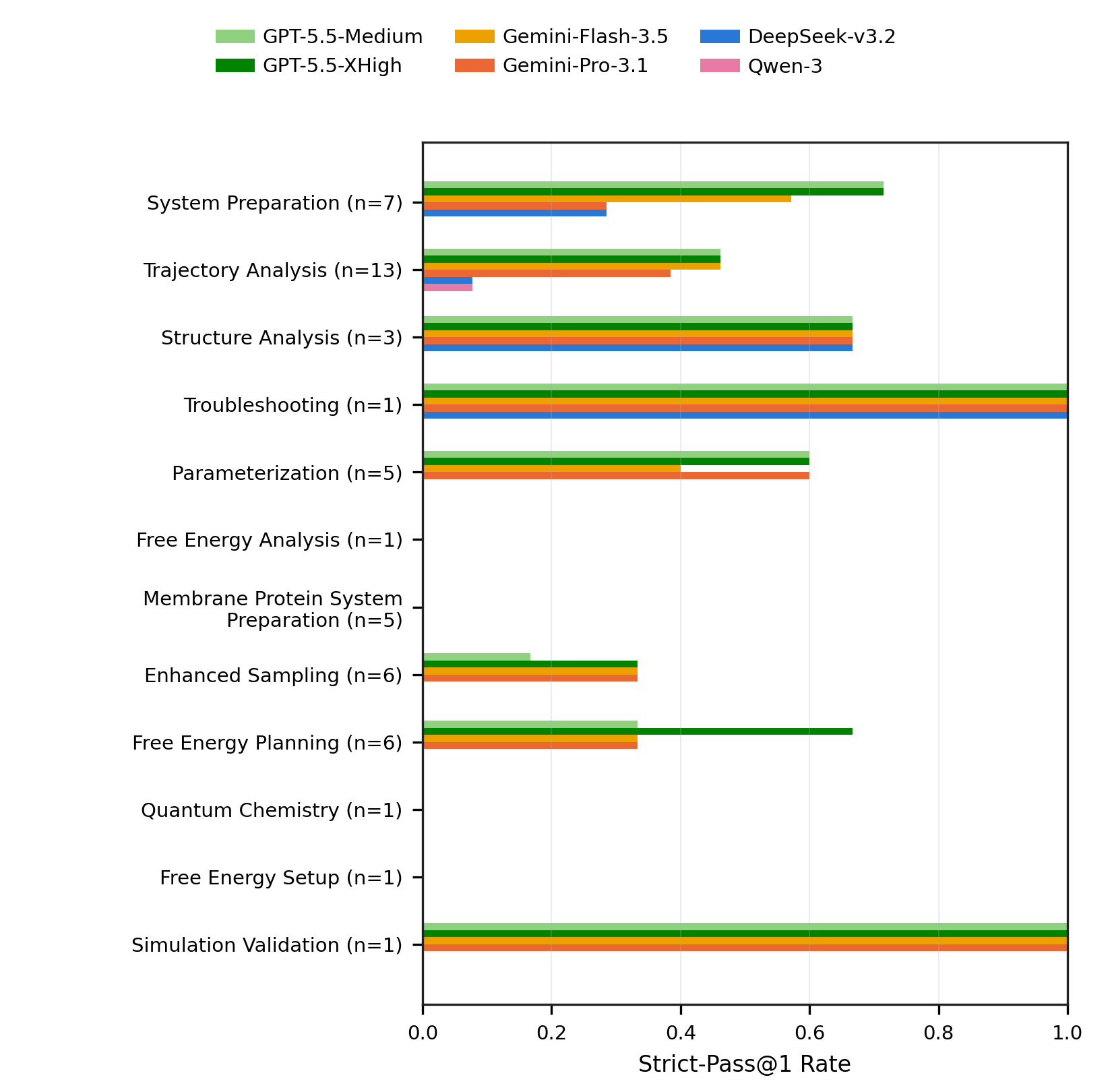}
\caption{Agent success rate by task category. Strict-Pass@1 success rate for each agent, broken down across MDArena task categories. Error bars show standard errors across tasks, using binomial standard errors for pass/fail success rates}
\label{fig:success-by-task-category}
\end{figure}

Performance varies sharply by task category (\cref{fig:success-by-task-category}). Agents are strongest on compact validation and troubleshooting-style tasks, though several of these categories have small sample sizes. System preparation, parameterization, and trajectory analysis are partly solved by the strongest agents, but are not fully solved by any evaluated configuration: the best-performing configurations solve 5/7 system preparation tasks (GPT-5.5 Medium and XHigh), 3/5 parameterization tasks (GPT-5.5 Medium, GPT-5.5 XHigh, Gemini Pro), and 6/13 trajectory analysis tasks (GPT-5.5 Medium, GPT-5.5 XHigh, Gemini Flash). Membrane-protein system preparation remains unsolved by every evaluated configuration. Higher-level tasks like free-energy planning further separate agents: GPT-5.5 XHigh solves 4/6 tasks in this category, while GPT-5.5 Medium, Gemini Flash 3.5, and Gemini Pro 3.1 each solve 2/6.

Correctness and process rewards, however, paint a somewhat different picture, with stronger models achieving noticeably higher scores even on tasks they ultimately fail to solve (see \cref{fig:category-check-scores}). These category-level results illustrate a broader pattern: as tasks grow more complex and multi-step, the likelihood that an agent overlooks a key step rises accordingly, and errors compound across steps, increasing the chance of eventual task failure even when individual steps are handled competently.

\subsection{Observations on agent performance}
Because MDArena tasks vary across systems, methods, packages, and instruction styles, it is impractical to isolate each dimension systematically. Instead, this section presents several representative case studies that capture broad, recurring trends that we consider most indicative of agent performance, focusing on dimensions along which success and failure patterns are particularly clear.

\subsubsection{Well-specified tasks are comparatively tractable} 
Well-specified tasks are comparatively tractable for AI agents. Even multi-step tasks are handled well when each step is defined in fine detail. While we deliberately left some hard tasks underspecified to probe agent judgment, we observed that certain easy and medium tasks, despite involving what would be a tedious search through a large parameter space for a human, were comparatively easy for agents. Several such tasks are solved by most or all evaluated configurations, including concentration/trajectory compatibility checks, alternate-location cleanup, clash checking, simple GROMACS parameter edits, and some ligand parameterization tasks. These tasks are compact, have clearly defined outputs, and typically admit direct validation with standard tools. They represent settings in which agents can already be useful under supervision.

\subsubsection{Long-horizon, high-level workflows remain brittle}
Long-horizon, high-level workflows, especially those with multiple valid solutions, remain brittle. The hardest tasks combine multiple packages and long dependency chains, such as membrane-protein system preparation, alchemical free-energy setup, and enhanced-sampling workflows. Failures typically emerge in the middle to late stages of an agent's trajectory but compound as they propagate through the remaining steps. A recurring driver of this failure mode is that agents tend to write bespoke scripts rather than rely on established packages such as MDAnalysis ~\citep{michaud2011mdanalysis}, even when those packages are already available in the containerized environment. Although quick to write, ad hoc agent-written scripts often overlook file-format conventions and edge cases handled by mature software. One representative example: an agent adding residues from a PDB file into a GRO file with its own script failed to account for differing residue-numbering conventions between the two formats. In isolation, this would be a minor and easily fixed error. Left unaddressed, however, it compounded into topology-coordinate mismatches later in the pipeline that agents consistently struggle to resolve. This illustrates how a reflexive preference for writing new code, rather than reusing robust, purpose-built tools, can backfire over the course of a long workflow.

\subsubsection{Overlooking fine details unrelated to the primary task}
A distinct failure mode, most pronounced in weaker models, involves overlooking task-relevant details that fall outside the immediate focus of the instruction. Higher-reasoning, stronger models tend to avoid this pitfall by checking their outputs extensively before concluding a task; weaker models instead declare success without robust verification. This can also be read as a manifestation of agents having a tendency to act without fully reasoning through why each step is necessary, itself a symptom of the underspecification issue discussed above. For example, Task 12 asks the agent to transfer crystal waters from a PDB file into a GRO file. Crystal waters in PDB files are conventionally represented by the oxygen atom alone; correctly completing the task requires isolating these waters, parameterizing them with a compatible force field, and transferring the resulting waters back with hydrogens included. Instead, agents frequently transfer only the oxygen atoms and report the task complete.

Qwen exhibits this failure mode heavily on easy tasks, more so than on medium or harder tasks, its failures are more fundamental, typically reflecting misinterpretation of the task or breakdowns in planning rather than oversight. On easy tasks, it typically completes the core objective but overlooks a peripheral detail: in a topology preparation task, for instance, Qwen correctly stages the resulting .top and .gro files into the output directory but omits the associated .itp files referenced by the topology, causing the correctness check to fail. This illustrates that even nominally successful task completion can fail on details incidental to, but required by, the stated objective, underscoring the gap between an agent appearing to finish a task and actually satisfying its full specification.

\subsubsection{Operational stability varies across agents and harnesses}

Operational failures were also strongly agent and harness dependent. Across 300 runs, 67 ended with an exception rather than a normal, verifier-scored outcome. Timeouts were the dominant failure mode, accounting for 36/67 errors overall, and were especially concentrated in Gemini Pro 3.1, where all 9 errored runs were timeouts. Qwen3-235B was the least operationally stable configuration, with 32/50 runs ending in error (\cref{fig:outcome-stacked-by-model}); its dominant failure mode was a nonzero agent exit, typically caused by the provider's context-limit failures rather than a genuine verifier failure. By contrast, the Codex configurations were operationally stable throughout: GPT-5.5 XHigh produced only 1 errored run, and GPT-5.5 Medium just 4. Taken together, these patterns suggest operational instability is often a harness-model coupling issue rather than a purely intrinsic model weakness: models that consume more context, through verbose tool calls or extended reasoning, are more likely to exceed time or context budgets regardless of underlying reasoning quality.

\section{Discussion}

Performance on MDArena shows that while coding agents can already be useful within clearly bounded task chunks, they remain far from autonomous computational researchers. Even the strongest evaluated configuration, GPT-5.5 XHigh, solves only 24/50 tasks (48\%) at Strict-Pass@1, despite comparatively strong average correctness (0.838) and process (0.856) scores, and near-perfect operational stability (1/50 error runs). This gap between high partial credit and modest full success is the frontier for the next generation of coding agents: while some of it may be closed with better tooling and domain-specific skills, we believe true generalizable improvement will require gains in raw model intelligence. The evaluated open-weight models, meanwhile, remain far from competitive on this benchmark, and closing that gap will likely require substantial further progress before they are viable for MD research use.

MDArena has several limitations. Each model-harness pair was evaluated only once per task, so our results reflect a practical baseline rather than repeated-run reliability, and stochastic variation across runs remains uncharacterized. The process verifier relies on an LLM judge, which, despite criterion-anchored rubrics, introduces its own model dependence. The task distribution is drawn from a single research group and may reflect that group's particular expertise and conventions rather than the full diversity of MD practice. Because task selection deliberately emphasized workflows expected to expose known agent failure modes, the resulting success rates should not be interpreted as estimates of performance over all routine MD tasks. Finally, the benchmark intentionally emphasizes system setup and analysis over long-running, stochastic production simulation; extending MDArena to evaluate end-to-end simulation execution is an important direction for future work, provided this is done without collapsing into the trap of grading stochastic simulation output directly.

Importantly, MDArena is intended as an extensible, longitudinal evaluation framework rather than a fixed snapshot. The initial task suite provides a versioned reference point, while future releases can incorporate additional systems, methods, and community-contributed workflows. Evaluations of recently released frontier models (e.g., GPT 5.6-Sol, Claude Opus-5, and Gemini-3.6-Flash) are already underway and will be soon incorporated into the next versioned release. More broadly, applying a consistent evaluation protocol across successive model–harness generations will allow MDArena to track both overall capability gains and persistent category-specific weaknesses, thereby meaningfully guiding the development of future AI agents for biomolecular simulations.

\section*{Data and Code Availability}
The MDArena benchmark, including task definitions and inputs, Docker specifications, verifier implementations, reference solutions, and evaluation configurations, is publicly available at \href{https://github.com/weitse-hsu/MDArena}{https://github.com/weitse-hsu/MDArena}. The version corresponding to the results reported in this manuscript is archived as v0.1.

\section*{Acknowledgements}

NMA is funded by the MRC via an iCASE studentship with Vertex Pharmaceuticals Europe Ltd with contributions from Reuben College, Oxford.  We thank Fabio Zuccotto, Ewa Chudyk and Ron Knegtel for useful discussions. We also acknowledge the BBSRC (JDC, PCB) and the Wellcome Trust (JDC, PCB).  EJG is supported by a Watson Scholarship and Corpus Christi College.  MS is in receipt of a  Department of Biochemistry scholarship.  KV acknowledges the Scripps/Oxford programme.  W-TH and LXP thank Ineos for support.

\bibliography{paper}
\bibliographystyle{iclr2021_conference}

\appendix
\section{Appendix}
\renewcommand{\thefigure}{A\arabic{figure}}
\renewcommand{\thetable}{A\arabic{table}}
\setcounter{figure}{0}
\setcounter{table}{0}

\subsection{Task Metadata Tags}
\cref{tab:task-metadata} lists the task-level metadata used to summarize MDArena's benchmark coverage. Package, system, and method cells contain semicolon-separated tags when a task belongs to multiple categories; blank package entries indicate that no explicit package/tool tag was assigned in the source package table.

\begingroup
\scriptsize
\setlength{\tabcolsep}{2pt}
\renewcommand{\arraystretch}{1.12}
\begin{longtable}{@{}L{0.18\textwidth}L{0.25\textwidth}L{0.24\textwidth}L{0.29\textwidth}@{}}
\caption{Task metadata tags for MDArena.}\label{tab:task-metadata}\\
\toprule
Task & Package tags & System tags & Method tags \\
\midrule
\endfirsthead
\caption[]{Task metadata tags for MDArena (continued).}\\
\toprule
Task & Package tags & System tags & Method tags \\
\midrule
\endhead
\midrule
\multicolumn{4}{r}{Continued on next page}\\
\endfoot
\bottomrule
\endlastfoot
\texttt{01\_\allowbreak{}similes2sim} & GROMACS / gmx & Generic solvated ligand from SMILES & System preparation / model building \\
\texttt{02\_\allowbreak{}conc\_\allowbreak{}diff\_\allowbreak{}sys\_\allowbreak{}traj} & GROMACS / gmx & Chignolin; \allowbreak{}T4 lysozyme & Simulation compatibility / sanity checks \\
\texttt{03\_\allowbreak{}id\_\allowbreak{}bp\_\allowbreak{}resid} & -- & Adenosine A2A receptor / A2AR & Structure analysis / validation \\
\texttt{04\_\allowbreak{}traj\_\allowbreak{}ana\_\allowbreak{}dist} & -- & Unspecified residue-distance trajectory system & Trajectory analysis \\
\texttt{05\_\allowbreak{}traj\_\allowbreak{}ana\_\allowbreak{}hbonds} & MDAnalysis & Unspecified protein-ligand H-bond trajectory & Trajectory analysis \\
\texttt{06\_\allowbreak{}traj\_\allowbreak{}ana\_\allowbreak{}rotamers} & -- & p38 alpha & Trajectory analysis \\
\texttt{07\_\allowbreak{}tyk2\_\allowbreak{}equil} & GROMACS / gmx & TYK2 & System preparation / model building \\
\texttt{08\_\allowbreak{}clash\_\allowbreak{}check} & -- & TYK2 & Structure analysis / validation; \allowbreak{}Simulation compatibility / sanity checks \\
\texttt{09\_\allowbreak{}holo2apo} & GROMACS / gmx & mGlu5 receptor & System preparation / model building \\
\texttt{10\_\allowbreak{}traj\_\allowbreak{}ana\_\allowbreak{}mg\_\allowbreak{}coord} & -- & Isocitrate lyase / ICL family & Trajectory analysis \\
\texttt{11\_\allowbreak{}amb2gmx\_\allowbreak{}em} & GROMACS / gmx & Glycosylated insulin / glycoinsulin & System preparation / model building \\
\texttt{12\_\allowbreak{}xtal\_\allowbreak{}waters} & MDAnalysis & Isocitrate lyase / ICL family & Structure analysis / validation \\
\texttt{13\_\allowbreak{}altloc\_\allowbreak{}clean} & -- & Oxymyoglobin & System preparation / model building \\
\texttt{14\_\allowbreak{}traj\_\allowbreak{}ana\_\allowbreak{}lig\_\allowbreak{}rmsd} & MDAnalysis & SLCO2A1 & Trajectory analysis \\
\texttt{15\_\allowbreak{}grompp\_\allowbreak{}tcoup} & GROMACS / gmx & Generic GROMACS protein system & Topology / force-field editing; \allowbreak{}Simulation compatibility / sanity checks \\
\texttt{16\_\allowbreak{}gaff\_\allowbreak{}param} & GROMACS / gmx; \allowbreak{}ACPYPE & Ala-Phe ligand & Ligand / residue parameterization \\
\texttt{17\_\allowbreak{}prolif\_\allowbreak{}int} & -- & Isocitrate lyase / ICL family & Trajectory analysis \\
\texttt{18\_\allowbreak{}scale\_\allowbreak{}charge} & GROMACS / gmx & Water-decane ion topology & Topology / force-field editing \\
\texttt{19\_\allowbreak{}traj\_\allowbreak{}ana\_\allowbreak{}open} & -- & TPC2 membrane protein & Trajectory analysis \\
\texttt{20\_\allowbreak{}fep\_\allowbreak{}abfe\_\allowbreak{}dhdl\_\allowbreak{}analysis} & alchemlyb & Adenosine A2A receptor / A2AR & FEP / alchemical free-energy setup \\
\texttt{21\_\allowbreak{}water\_\allowbreak{}transfer\_\allowbreak{}ox2r} & MDAnalysis & OX2R receptor & Structure analysis / validation \\
\texttt{22\_\allowbreak{}traj\_\allowbreak{}ana\_\allowbreak{}pept2\_\allowbreak{}pca\_\allowbreak{}cv\_\allowbreak{}suitability} & PLUMED & PepT2 / beta-lactam systems & Collective-variable analysis / PCA \\
\texttt{23\_\allowbreak{}a2ar\_\allowbreak{}membrane\_\allowbreak{}equilibration} & GROMACS / gmx; \allowbreak{}OpenMM; \allowbreak{}ACPYPE; \allowbreak{}Open Babel; \allowbreak{}PDBFixer & Adenosine A2A receptor / A2AR & System preparation / model building \\
\texttt{24\_\allowbreak{}ligand\_\allowbreak{}parameterization\_\allowbreak{}smirnoff\_\allowbreak{}openfe} & OpenFF Toolkit / SMIRNOFF workflow; \allowbreak{}OpenMM; \allowbreak{}openmmforcefields; \allowbreak{}AmberTools; \allowbreak{}ParmEd & Generic SMIRNOFF/OpenFF ligand parameterization & Ligand / residue parameterization \\
\texttt{25\_\allowbreak{}mglu5\_\allowbreak{}membrane\_\allowbreak{}setup} & GROMACS / gmx; \allowbreak{}OpenFF Toolkit / SMIRNOFF workflow; \allowbreak{}OpenMM; \allowbreak{}openmmforcefields; \allowbreak{}AmberTools; \allowbreak{}ParmEd; \allowbreak{}PDBFixer & mGlu5 receptor & System preparation / model building; \allowbreak{}Ligand / residue parameterization \\
\texttt{26\_\allowbreak{}d3\_\allowbreak{}receptor\_\allowbreak{}membrane\_\allowbreak{}setup} & GROMACS / gmx; \allowbreak{}MDAnalysis; \allowbreak{}OpenMM; \allowbreak{}ACPYPE; \allowbreak{}Open Babel; \allowbreak{}PDBFixer & Dopamine D3 receptor & System preparation / model building; \allowbreak{}Ligand / residue parameterization \\
\texttt{27\_\allowbreak{}a2ar\_\allowbreak{}6gt3\_\allowbreak{}md\_\allowbreak{}setup} & GROMACS / gmx; \allowbreak{}OpenMM; \allowbreak{}ACPYPE; \allowbreak{}Open Babel; \allowbreak{}PDBFixer & Adenosine A2A receptor / A2AR & System preparation / model building; \allowbreak{}Ligand / residue parameterization \\
\texttt{28\_\allowbreak{}traj\_\allowbreak{}ana\_\allowbreak{}pept2\_\allowbreak{}pca\_\allowbreak{}plumed\_\allowbreak{}setup} & PLUMED & PepT2 / beta-lactam systems & Collective-variable analysis / PCA; \allowbreak{}Umbrella sampling setup \\
\texttt{29\_\allowbreak{}pept2\_\allowbreak{}beta\_\allowbreak{}lactam\_\allowbreak{}membrane\_\allowbreak{}setup} & GROMACS / gmx; \allowbreak{}ACPYPE; \allowbreak{}Open Babel & PepT2 / beta-lactam systems & System preparation / model building; \allowbreak{}Ligand / residue parameterization \\
\texttt{30\_\allowbreak{}cam\_\allowbreak{}umbrella\_\allowbreak{}pore\_\allowbreak{}6nq0} & GROMACS / gmx; \allowbreak{}MDAnalysis & 6NQ0 pore / ion-conductance system & Umbrella sampling setup \\
\texttt{31\_\allowbreak{}traj\_\allowbreak{}ana\_\allowbreak{}na\_\allowbreak{}conductance\_\allowbreak{}6nq0} & GROMACS / gmx & 6NQ0 pore / ion-conductance system & Trajectory analysis \\
\texttt{32\_\allowbreak{}fep\_\allowbreak{}openfe\_\allowbreak{}rbfe\_\allowbreak{}dhodh} & OpenFE; \allowbreak{}OpenFF Toolkit / SMIRNOFF workflow; \allowbreak{}OpenMM; \allowbreak{}openmmtools; \allowbreak{}AmberTools; \allowbreak{}ParmEd; \allowbreak{}RDKit & DHODH & FEP / alchemical free-energy setup \\
\texttt{33\_\allowbreak{}fep\_\allowbreak{}openfe\_\allowbreak{}rbfe\_\allowbreak{}d3} & OpenFE; \allowbreak{}OpenFF Toolkit / SMIRNOFF workflow; \allowbreak{}OpenMM; \allowbreak{}openmmtools; \allowbreak{}AmberTools; \allowbreak{}ParmEd; \allowbreak{}RDKit & Dopamine D3 receptor & FEP / alchemical free-energy setup \\
\texttt{34\_\allowbreak{}fep\_\allowbreak{}openfe\_\allowbreak{}rbfe\_\allowbreak{}def\_\allowbreak{}ox2\_\allowbreak{}s1} & OpenFE; \allowbreak{}OpenFF Toolkit / SMIRNOFF workflow; \allowbreak{}OpenMM; \allowbreak{}openmmtools; \allowbreak{}AmberTools; \allowbreak{}ParmEd; \allowbreak{}RDKit & OX2R receptor & FEP / alchemical free-energy setup \\
\texttt{35\_\allowbreak{}fep\_\allowbreak{}openfe\_\allowbreak{}abfe\_\allowbreak{}d3} & OpenFE; \allowbreak{}OpenFF Toolkit / SMIRNOFF workflow; \allowbreak{}OpenMM; \allowbreak{}openmmtools; \allowbreak{}AmberTools; \allowbreak{}ParmEd; \allowbreak{}RDKit & Dopamine D3 receptor & FEP / alchemical free-energy setup \\
\texttt{36\_\allowbreak{}fep\_\allowbreak{}openfe\_\allowbreak{}abfe\_\allowbreak{}mglu5r} & OpenFE; \allowbreak{}OpenFF Toolkit / SMIRNOFF workflow; \allowbreak{}OpenMM; \allowbreak{}openmmtools; \allowbreak{}AmberTools; \allowbreak{}ParmEd; \allowbreak{}RDKit & mGlu5 receptor & FEP / alchemical free-energy setup \\
\texttt{37\_\allowbreak{}fep\_\allowbreak{}openfe\_\allowbreak{}abfe\_\allowbreak{}a2ar} & MDAnalysis; \allowbreak{}OpenFE; \allowbreak{}OpenFF Toolkit / SMIRNOFF workflow; \allowbreak{}OpenMM; \allowbreak{}openmmtools; \allowbreak{}AmberTools; \allowbreak{}ParmEd; \allowbreak{}RDKit & Adenosine A2A receptor / A2AR & FEP / alchemical free-energy setup \\
\texttt{38\_\allowbreak{}pept2\_\allowbreak{}qm\_\allowbreak{}cluster\_\allowbreak{}orca} & MDAnalysis; \allowbreak{}ORCA; \allowbreak{}MPI / OpenMPI / mpi4py & PepT2 / beta-lactam systems & QM / quantum chemistry setup \\
\texttt{39\_\allowbreak{}fep\_\allowbreak{}gmx\_\allowbreak{}d3\_\allowbreak{}abfe} & GROMACS / gmx; \allowbreak{}MDAnalysis; \allowbreak{}MDRestraintsGenerator & Dopamine D3 receptor & FEP / alchemical free-energy setup; \allowbreak{}Boresch restraints \\
\texttt{40\_\allowbreak{}traj\_\allowbreak{}ana\_\allowbreak{}meng\_\allowbreak{}sam\_\allowbreak{}rmsd} & -- & MenG membrane/SAM system & Trajectory analysis \\
\texttt{41\_\allowbreak{}rexee\_\allowbreak{}argon\_\allowbreak{}hydration\_\allowbreak{}setup} & GROMACS / gmx; \allowbreak{}ensemble\_md; \allowbreak{}MPI / OpenMPI / mpi4py & Argon hydration & REXEE \\
\texttt{42\_\allowbreak{}expanded\_\allowbreak{}ensemble\_\allowbreak{}mdp} & GROMACS / gmx & Generic expanded-ensemble MDP system & Topology / force-field editing; \allowbreak{}Expanded ensemble / Wang-Landau \\
\texttt{43\_\allowbreak{}cb7\_\allowbreak{}10\_\allowbreak{}weight\_\allowbreak{}updating\_\allowbreak{}ee} & GROMACS / gmx & CB7-10 host-guest complex & Expanded ensemble / Wang-Landau \\
\texttt{44\_\allowbreak{}traj\_\allowbreak{}ana\_\allowbreak{}atp8b\_\allowbreak{}rmsd} & MDAnalysis & ATP8B1-CDC50A membrane complex & Trajectory analysis \\
\texttt{45\_\allowbreak{}atp8b\_\allowbreak{}insertion\_\allowbreak{}depth} & -- & ATP8B1-CDC50A membrane complex & Trajectory analysis \\
\texttt{46\_\allowbreak{}meng\_\allowbreak{}dmk\_\allowbreak{}ff\_\allowbreak{}patch} & GROMACS / gmx & MenG membrane/SAM system & Topology / force-field editing \\
\texttt{47\_\allowbreak{}d2p\_\allowbreak{}residue\_\allowbreak{}parameterization} & GROMACS / gmx & Dianionic phosphoaspartate / D2P pentapeptide & Ligand / residue parameterization \\
\texttt{48\_\allowbreak{}cb7\_\allowbreak{}10\_\allowbreak{}alchemical\_\allowbreak{}metadynamics} & GROMACS / gmx; \allowbreak{}PLUMED & CB7-10 host-guest complex & Alchemical metadynamics \\
\texttt{49\_\allowbreak{}slco2a1\_\allowbreak{}physical\_\allowbreak{}validation} & GROMACS / gmx; \allowbreak{}physical\_validation & SLCO2A1 & Structure analysis / validation; \allowbreak{}Simulation compatibility / sanity checks \\
\texttt{50\_\allowbreak{}glycoinsulin\_\allowbreak{}beta\_\allowbreak{}sheet\_\allowbreak{}propensity} & MDAnalysis; \allowbreak{}Matplotlib & Glycosylated insulin / glycoinsulin & Trajectory analysis \\
\end{longtable}

\endgroup

\subsection{Additional Outcome Breakdown}

\begin{figure}[!ht]
\centering
\includegraphics[width=\textwidth]{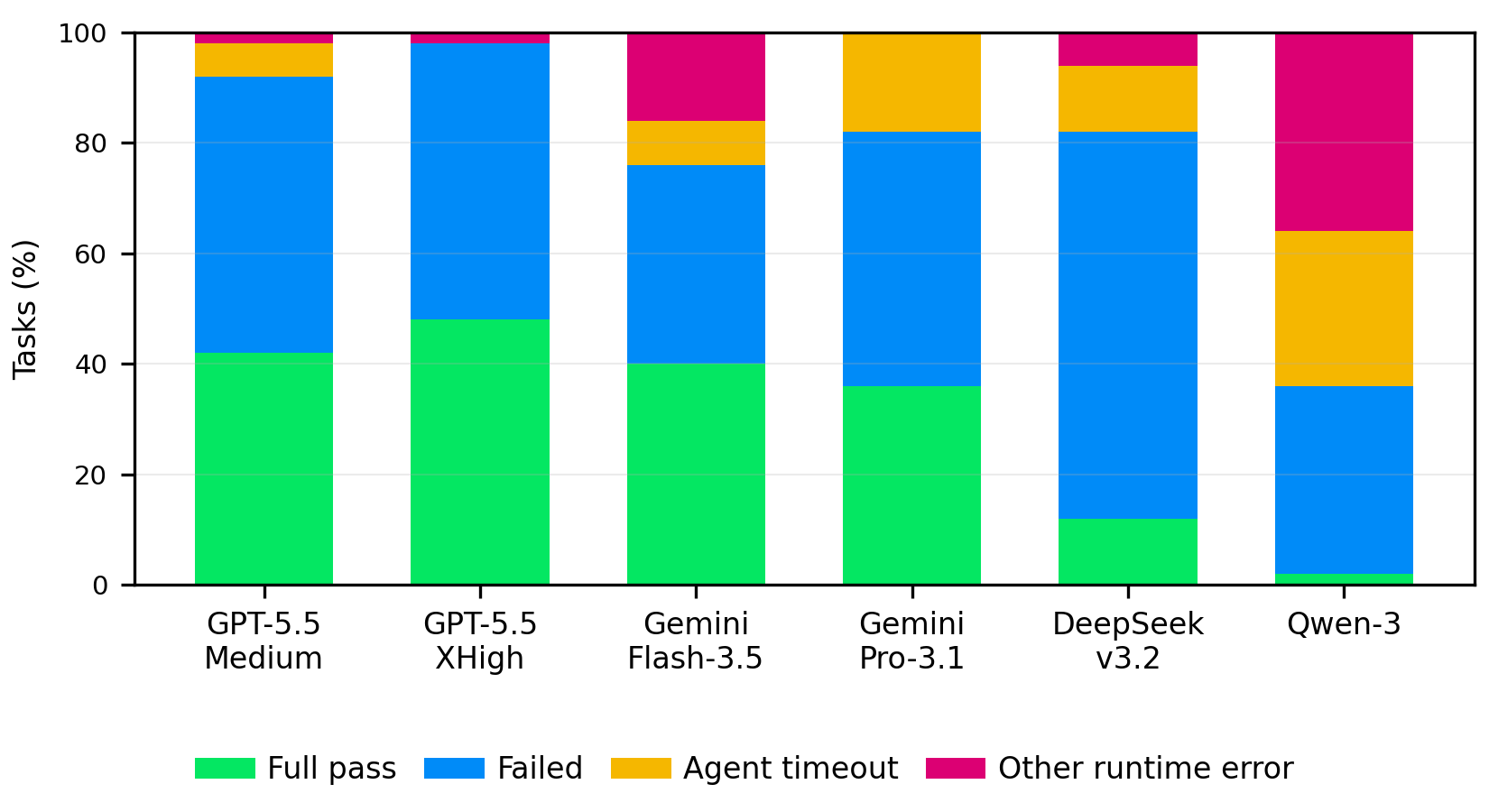}
\caption{Task outcomes by model. Stacked bars show the percentage of MDArena tasks ending in a full pass, failed evaluation, agent timeout, or other runtime error.}
\label{fig:outcome-stacked-by-model}
\end{figure}

\subsection{Category-Wise Check Scores}
\begin{figure}[!ht]
\centering
\includegraphics[width=\textwidth]{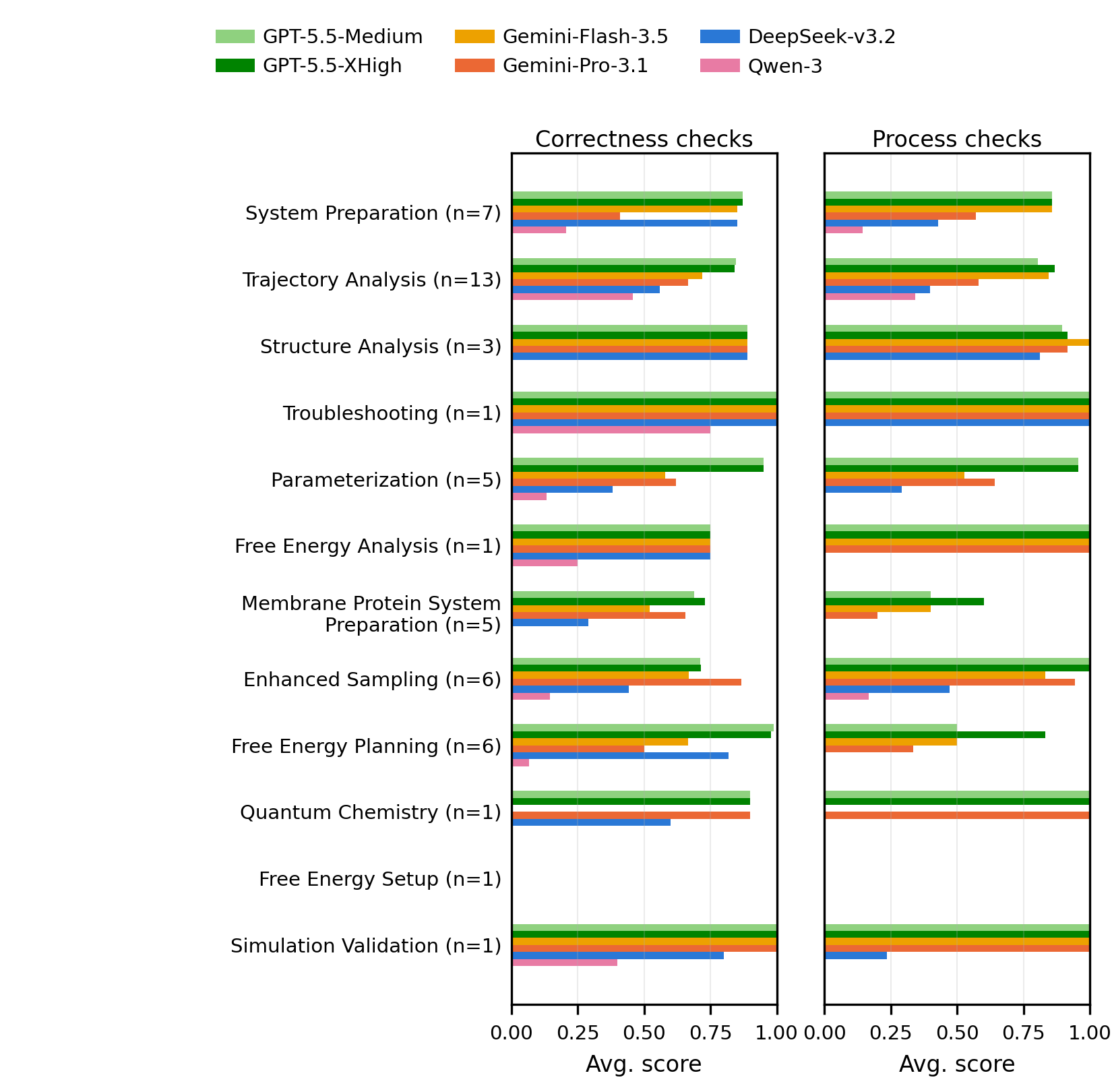}
\caption{Category-wise correctness and process rewards. Average check-level reward by task category for each model/harness configuration, separated into (a) correctness checks and (b) process checks.}
\label{fig:category-check-scores}
\end{figure}

\end{document}